\documentclass[10pt,twocolumn,twoside] {IEEEtran}

\usepackage{amsmath,graphicx}
\usepackage{amsthm}
\usepackage{amscd}
\usepackage{amssymb}
\usepackage{latexsym}
\usepackage{algorithm}
\usepackage{algpseudocode}
\usepackage{cite}
\usepackage{multirow}
\usepackage{tikz}
\usepackage{epstopdf}
\usepackage{url}
\ifCLASSOPTIONcompsoc
\usepackage[caption=false,font=normalsize,labelfont=sf,textfont=sf]{subfig}
\else
\usepackage[caption=false,font=footnotesize]{subfig}
\fi

\newtheorem{theorem}{Theorem}[section]
\newtheorem{lemma}[theorem]{Lemma}

\newtheorem{dfn}[theorem]{Definition}
\newtheorem{remark}[theorem]{Remark}

\newcommand{\argmin}{\text{argmin}}

\newcommand{\overbar}[1]{\mkern 1.5mu\overline{\mkern-1.5mu#1\mkern-1.5mu}\mkern 1.5mu}

\def\x{\mathbf{x}}
\def\xh{\widehat{\mathbf{x}}}

\def\om0{\omega}

\def\y{\mathbf{y}}

\def\bigo{\mathcal{O}}
\def\a{\mathbf{a}}

\def\s{\mathbf{s}}
\def\sh{\hat{\mathbf{s}}}

\def\w{\mathbf{w}}

\def\A{\mathbf{A}}
\def\x{\mathbf{x}}
\def\V{\mathbf{V}}
\def\U{\mathbf{U}}

\def\P{\mathbb{P}}
\def\E{\mathbb{E}}

\def\Z{\mathbf{Z}}
\def\N{\mathbb{N}}

\def\u{\mathbf{u}}
\def\n{\mathbf{n}}

\def\W{\mathbf{W}}

\def\pphi{\boldsymbol{\phi}}

\def\PPhi{\boldsymbol{\Phi}}
\def\H{\mathbf{H}}
\def\I{\mathbf{I}}
\def\h{\mathbf{h}}
\def\G{\mathbf{G}}
\def\C{\mathbb{C}}
\def\Y{\mathbf{Y}}
\def\N{\mathbf{N}}
\def\X{\mathbf{X}}
\def\Xb{\overbar{\mathbf{X}}}

\def\tr{\mathrm{tr}}
\def\totP{\mathcal{P}}
\def\Hls{\widehat{\mathbf{H}}_{LS}}
\def\HLSsls{\widehat{\mathbf{H}}_{LS-SLS}}
\def\Hmmse{\widehat{\mathbf{H}}_{MMSE}}

\def\Had{\widehat{\mathbf{H}}_{AD}}

\def\Hh{\widehat{\mathbf{H}}}
\def\Wh{\widehat{\mathbf{W}}}

\def\mA{\mathcal{A}}
\def\s{\mathbf{s}}
\def\srho{\sqrt{\rho}}
\def\minfty{\overset{M\rightarrow\infty}{\rightarrow}}
\def\Ab{\overbar{\mathbf{A}}}
\def\yb{\overbar{\mathbf{y}}}
\def\nb{\overbar{\mathbf{n}}}
\def\Hb{\overbar{\mathbf{H}}}
\def\xb{\overbar{\mathbf{x}}}
\def\ab{\overbar{\mathbf{a}}}
\def\mnorm{2,1}
\def\minfnorm{2,\infty}

\def\B{\mathbf{B}}
\def\P{\mathbf{P}}
\def\Q{\mathbf{Q}}
\def\F{\mathbf{F}}
\def\G{\mathbf{G}}
\def\Tr{\text{Tr}}

\newcounter{MYtempeqncnt}

\begin{document}

\title{Atomic Norm Denoising-Based Joint Channel Estimation and Faulty Antenna Detection for Massive MIMO}

\author{Peng~Zhang, Lu~Gan, Cong~Ling and~Sumei~Sun
\thanks{The work was supported by the Agency for Science, Technology and Research (A$^*$STAR) through the AGS(O) scholarship.}
\thanks{P. Zhang is with the Institute for Infocomm Research, A$^*$STAR, Singapore 138632, and also with the Department of Electrical and Electronic Engineering, Imperial College London, London, SW7 2AZ, UK (e-mail: zhangp@i2r.a-star.edu.sg; p.zhang12@imperial.ac.uk).}%
\thanks{L. Gan is with the College of Engineering, Design and Physical Science, Brunel University, London, UB8 3PH, UK (e-mail: lu.gan@brunel.ac.uk).}
\thanks{C. Ling is with the Department of Electrical and Electronic Engineering, Imperial College London, London,
SW7 2AZ, UK (e-mail: cling@ieee.org) .}
\thanks{S. Sun is with the Institute for Infocomm Research, A$^{*}$STAR, Singapore, 138632, Singapore (e-mail: sunsm@i2r.a-star.edu.sg).}}




\maketitle

\begin{abstract}
We consider joint channel estimation and faulty antenna detection for massive multiple-input multiple-output (MIMO) systems operating in time-division duplexing (TDD) mode. For systems with faulty antennas, we show that the impact of faulty antennas on uplink (UL) data transmission does not vanish even with unlimited number of antennas. However, the signal detection performance can be improved with a priori knowledge on the indices of faulty antennas. This motivates us to propose the approach for simultaneous channel estimation and faulty antenna detection. By exploiting the fact that the degrees of freedom of the physical channel matrix are smaller than the number of free parameters, the channel estimation and faulty antenna detection can be formulated as an extended atomic norm denoising problem and solved efficiently via the alternating direction method of multipliers (ADMM). Furthermore, we improve the computational efficiency by proposing a fast algorithm and show that it is a good approximation of the corresponding extended atomic norm minimization method. Numerical simulations are provided to compare the performances of the proposed algorithms with several existing approaches and demonstrate the performance gains of detecting the indices of faulty antennas.
\end{abstract}


\section{Introduction}
Massive multiple-input multiple-output (MIMO) or large antenna systems have been proposed in \cite{marzetta2010noncooperative}, where each base station (BS) is equipped with a large number of antennas that greatly exceeds the number of user terminals (UTs). As a potential enabler for the development of future broadband network, it offers many advantages over conventional point-to-point MIMO, such as energy-efficiency, security, and robustness \cite{larrson14massiv}. However, the spatial multiplexing gains and the array gains of massive MIMO rely on accurate channel state information at the transmitter (CSIT) \cite{larrson14massiv,ngo2013energy}.

In time-division duplexing (TDD) massive MIMO systems, only CSI for the uplink needs to be estimated due to the assumption of channel reciprocity. Conventional methods on massive MIMO channel estimation include least squares (LS) and minimum mean squared error (MMSE) \cite{hoydis2013massive,rusek2013scaling,ngo2011analysis}. A compressed sensing (CS) based approach has been proposed to address the channel estimation for massive MIMO systems with a physical finite scattering channel model \cite{nguyen2013compressive}. By employing random Bernoulli pilots at each user terminal (UT), the channel estimation was converted into a low-rank approximation problem and solved via quadratic semidefinite programming (SDP). Recently, a novel atomic norm denoising-based channel estimation method was proposed to further improve the performance \cite{peng15atomic}. In \cite{peng15atomic}, it is demonstrated that the atomic norm denoising-based approach can outperforms the least squares, the least squares scaled least squares \cite{biguesh2006training} and the compressed sensing-based methods \cite{nguyen2013compressive}\cite{rao2014distributed}. Similar atomic norm denoising-based channel estimation method has also been used for OFDM systems in \cite{pejoski2015estimation,pejoski2016joint}.

However, an important issue overlooked by research on channel estimation for massive MIMO is the impact of faulty antennas at the BS. The use of large-scale antenna arrays brings many remarkable features, such as propagation loss mitigation and low inter-user interference \cite{marzetta2010noncooperative,larrson14massiv,ngo2013energy}. For the consideration of network deployment, it might be more attractive and practical to use large-scale antenna arrays with inexpensive antenna elements. Thus, there is always a possibility of failure of one or more elements in the large array \cite{peters1991conjugate,yeo1999array,acharya2011null,de2013massive}. It is therefore important to analyze the impact of faulty antennas on the channel estimation accuracy and data transmission quality. Moreover, it would be significant to promptly identify the faulty antennas, which is crucial for replacement or reconfiguration. We note that the influence of hardware impairments (e.g., amplifier nonlinearities, I/Q imbalances, and quantization errors) has been considered in  \cite{bjornson2013capacity,koch2009channels,studer2013aware,mohammed2013per,bjornson2014massive}. Specifically, \cite{bjornson2014massive} analyzed the aggregate impacts of different hardware impairments on massive MIMO systems. However, the problem of hardware impairments is different from the impact of faulty antennas: the distortion noise caused by the hardware impairments is modeled as additive noise (e.g., Gaussian distribution conditioned on given channel realization) and applied to all the antennas at the BS \cite{schenk2008rf,studer2010mimo,bjornson2014massive}; on the other hand, the corruption noise due to faulty antennas usually appears only on a small subset of the antenna array, but with unknown distribution and unbounded magnitudes (see Section \ref{sec: sys and chn model} for the system model).

In this paper, we consider the channel estimation for massive MIMO systems under the impact of faulty antennas. We analyze the mean squared error of the signal detection with maximum ratio combining (MRC) and zero-forcing (ZF) receivers on the uplink (UL) in the large system limit. It has been shown in \cite{marzetta2010noncooperative} that a linear receiver can be used to eliminate the effect of uncorrelated noise and small-scale fading when considering an infinite number of fault-free antennas. Here, we prove that the effect of faulty antennas does not vanish even with unlimited number of antennas. Furthermore, we show that the knowledge on the indices of faulty antennas can be used to modify the receiver for better signal detection performance. In particular, the resultant MSE is always smaller than the original one in the large system limit. It is noted that the improvement on signal detection depends on the ability to acquire the indices of faulty antennas. Our main contribution is an atomic denoising-based method that can estimate the channel and detect faulty antennas simultaneously in the training phase. Relying on this advantage, the system will be able to: 1) monitor the antenna array; 2) identify the faulty antennas; and 3) maintain good signal detection performance before repair. In addition, we consider reducing the computational complexity of the proposed method for practical implementation. We propose a fast algorithm based on the group lasso and prove that it can achieve nearly the same performance as the atomic norm-based method.

{\emph{Additional related work.}} Detection of the faulty elements in large antenna arrays has been considered in \cite{rodriguez2009rapid,lee1988near,bucci2000diagnosis,migliore2011compressed,oliveri2012reliable}. In these works, different techniques are proposed to identity the faulty elements based on the measurements of the ``fault-free'' and ``distorted'' radiation patterns, e.g., back-propagation \cite{lee1988near}, genetic algorithms \cite{bucci2000diagnosis} and Bayesian compressed sensing \cite{oliveri2012reliable}. In contrast, our work focuses on simultaneous channel estimation and faulty antenna detection from the received signals in pilot-aided training. In addition, our atomic norm denoising-based approaches are novel in the faulty antennas detection literature. On the other hand, antenna selection has been studied to help to reduce the system complexity and cost by using a smaller number of RF chains than the antennas \cite{dua2006receive,mahboob2012transmit,gao2013antenna,gao2015massive}. This is different from our work since they assume that all the antennas are working and the CSI over all the antennas is available.

{\emph{Notations.}} We denote matrices and vectors by uppercase and lowercase boldface letters respectively. For a vector $\a$, we denote by $a_i$, ($i\in[n]=\{0,...,n-1\}$), its $i$-th element. We represent a sequence of vectors by $\a_0,...,\a_{n-1}$. For a matrix $\A$, $\A_{(j,k)}$ denotes the element on its $j$-th row and $k$-th column. The vector obtained by taking the $j$-th row ($k$-th column) of $\A$ is represented by $\A_{(j,:)}$ ($\A_{(:,k)}$). $\A^{-1}$, $\A^{\dagger}$ and $\A^*$ represent the inverse, the pseudo-inverse and the conjugate transpose of $\A$. The Frobenius norm and spectral norm of $\A$ is denoted by $\|\A\|_F=\sqrt{\text{tr}(\A^*\A)}$ and $\|\A\|_2$ respectively.

{\emph{Outline.}} The rest of the paper is organized as follows. In Section \ref{sec: sys and chn model}, we first introduce the system and channel models, then  review the concept of atomic norm and its application in channel estimation for faultless antennas. In Section \ref{sec: chn est with defective antennas}, we consider the channel estimation under faulty antennas. In particular, we analyse the impact of faulty antennas on UL data transmission, propose an approach to simultaneously estimate the channel and identify the faulty antennas, and provide a fast algorithm that can well approximate the atomic norm-based approach. Section \ref{sec: simulations} demonstrates the performance of the proposed approaches through a series of numerical simulations, followed by discussions and conclusions in Section \ref{sec: discussion} and \ref{sec: conclusion} respectively.

\section{Preliminaries}\label{sec: sys and chn model}
\subsection{Massive MIMO system model}
For a massive MIMO system operating in TDD mode, only CSI for the uplink needs to be estimated due to the assumption of channel reciprocity. There is one BS equipped with a large number of antennas $M$. In the uplink, $K$ autonomous single-antenna UTs send signals to the BS ($M\geq K$). At the $t$-th symbol time, the received signal $\y_t\in\C^M$ at the BS can be expressed as
\begin{align}\label{eqn: single symbol chn model}
\y_t=\H\x_t+\n_t,
\end{align}
where $\H\in\C^{M\times K}$ is the flat-fading quasi-static channel matrix, $\x_t\in\C^K$ denotes the transmitted vector of the $K$ users and $\n_t\in\C^M$ is the complex Gaussian noise with zero mean and unit variance. In the training phase, each user sends a training sequence of length $L$. The signal model \eqref{eqn: single symbol chn model} can be equivalently written as
\begin{align}\label{eqn: L symbols chn model}
\Y=\H\X+\N_0,
\end{align}
where $\X=[\x_1, \x_2, \dots,\x_L]\in\C^{K\times L}$ is the training matrix, $\Y\in\C^{M\times L}$ is the received signal matrix and $\N_0\in\C^{M\times L}$ is the noise matrix with independent and identically distributed (i.i.d.) $\mathcal{CN}(0,\sigma^2)$ entries. The training matrix $\X=[\pphi_1^T; \pphi_2^T; \dots; \pphi_K^T]$ can be regarded as the collection of $K$ length-$L$ training sequences from the UTs, where $\pphi_k$ denotes the training sequence transmitted from user $k$.

In this paper, we consider systems with faulty antennas at the BS. Let $S=\gamma M$, $\gamma\in[0,1]$, denote the number of faulty antennas, then at the $t$-th symbol time, the distortion caused by faulty antennas is denoted by an $S$-sparse vector $\w_t$, in which case the received signal becomes $\y_t=\H\x_t+\w_t+\n_t$. The support of $\w_t$'s indicates the indices of faulty antennas, which is assumed to be arbitrary and static for a relative long period. However, at different symbol times, the magnitude of the sparse vectors may vary arbitrarily. The signal model in the training phase is given by\begin{align}\label{eqn: L symbols chn model distortion}
\Y=\H\X+\W_0+\N_0,
\end{align}
where $\W_0\in\C^{M\times L}$ is a matrix with $S$ non-zero rows to account for the corruption caused by the $S$ faulty antennas. The indices of non-zero rows of $\W_0$ are equivalent to the positions of faulty antennas at the BS. Our main task is to estimate the channel matrix $\H$ and the positions of faulty antennas based on the training matrix $\X$ and the received signal matrix $\Y$. We note that the system model \eqref{eqn: L symbols chn model distortion} represents a general scenario, where no statistical knowledge on the channel matrix or the corruption is required.

\subsection{Finite scattering channel model}
In this paper, we consider a physical finite scattering model that has been frequently used in the literature for massive MIMO systems \cite{nguyen2013compressive,NHQ13Massive,tsai2002impact,ngo2011analysis,zuo2017energy,alodeh2015spatial}. This practical channel model is motivated by two reasons. First, it characterizes a poor scattering propagation environment, where the number of physical objects is limited \cite{muller2002random,burr2003capacity,fuhl1998unified,petrus2002geometrical}. Second, it can also describe the propagation channel where the scatters appear in groups (a.k.a. clusters) with similar delays, angle-of-arrivals (AoAs) and angle-of-departures (AoDs) \cite{rusek2013scaling, chizhik2002keyholes}. In this channel model, the angular domain is divided into a finite number of directions $P$, which is fixed and smaller than the number of BS antennas. For uniform linear arrays (ULA), the steering vector associated with each angle-of-arrival (AOA) $\theta_p\in[-\pi/2, \pi/2]$, $p=1,\dots,P$, is given by \cite{tsai2002impact}
\begin{align}\label{eqn: steering vector}
\a(\theta_p) \triangleq [1 \quad \mathrm{e}^{-j2\pi\frac{D}{\lambda}\sin(\theta_p)} \dots \quad \mathrm{e}^{-j2\pi\frac{(M-1)D}{\lambda}\sin(\theta_p)}]^T,
\end{align}
where $D$ is the antenna spacing at the BS and $\lambda$ is the signal wavelength. The channel vector from UT $k$ to the BS is defined as
\begin{align*}
\h_k=\frac{1}{\sqrt{P}}\sum_{p=1}^Pg_{kp}\a(\theta_p),
\end{align*}
where $g_{kp}$ is the random propagation coefficient from user $k$ to the BS. Hence, the channel matrix $\H$ can be written as
\begin{align*}
\H=\frac{1}{\sqrt{P}}\A\G,
\end{align*}
where $\A=[\a(\theta_1), \a(\theta_2), \cdots, \a(\theta_P)]\in\C^{M\times P}$ contains $P$ steering vectors and the path gain matrix $\G\in\C^{P\times K}$ is assumed to have independent entries with $\G_{(p,k)}\sim\mathcal{CN}(0,\delta_{pk}^2)$. The variance $\delta_{pk}^2$ is the channel's average attenuation including the path loss and shadowing effects \cite{rusek2013scaling,yin2013coordinated}. In the rest of this paper, we assume that $\delta_{pk}^2=1$ just for the simplicity. We note that the analysis can be easily extended to different values of $\delta_{pk}$ with more general distance-based power-law decaying path-loss distributions.

\subsection{LS and MMSE channel estimation}
The conventional pilot-aided channel estimation is accomplished by transmitting a training sequence with length $L\geq K$ from each user to the BS. Based on the received signal matrix $\Y$ and the training matrix $\X$ as in \eqref{eqn: L symbols chn model distortion}, the least squares (LS), LS scaled least squares (LS-SLS) channel estimates can be written as \cite{biguesh2006training}
\begin{align}
\Hls&=\Y\X^{\dagger}=\Y\X^*(\X\X^*)^{-1}\\
\HLSsls&=\frac{K\tr(\Y\Y^H)}{\totP(\sigma^2KM+\tr(\Y\Y^H))}\Y\X^*,
\end{align}
and the minimum mean-square-error (MMSE) channel estimate is given by \cite{NHQ13Massive}
\begin{align}
\Hmmse&=\frac{K}{\totP}\A(\A^*\A)^{-1}\A^*\Y\X^*,
\end{align}
where the total power is constrained by $\mathcal{P}=\|\X\|_F^2$. We note that the MMSE method assumes that the matrix $\A$ is available, whereas our proposed channel estimation and faulty antenna detection schemes do not require such assumption.

\subsection{Atomic norm}
Suppose a high-dimensional signal $\x$ can be formed as \cite{chandrasekaran2012convex}
\begin{align*}
\x=\sum_{i=1}^rc_i\a_i, \quad \a_i\in\mA, c_r\geq 0,
\end{align*}
where $\mA$ is a set of atoms that consists of simple building blocks of general signals (a.k.a atomic set). $\x$ is said to be simple when it can be expressed as sparse linear combination of atoms from some atomic set (i.e., $r$ is relatively small). The atomic set $\mA$ can be very general: for example, if $\x$ is a sparse vector, $\mA$ could be the finite set of unit-norm one-sparse vectors; if $\x$ is a low-rank matrix, $\mA$ could be the set of all unit-norm rank-$1$ matrices. The atomic norm is defined as below.
\begin{dfn}[Atomic Norm]\cite{chandrasekaran2012convex}
The atomic norm $\|\cdot\|_{\mA}$ of $\mA$ is the Minkowski functional (or the gauge function) associated with the convex hull of $\mA$ ($\text{\emph{conv}}(\mA)$):
\begin{align}
\|\x\|_{\mA}=\inf\{t>0|\x\in t\text{\emph{conv}}(\mA)\}.
\end{align}
\end{dfn}

For example, when $\mA$ is the set of unit-norm one-sparse vectors in $\C^n$, the atomic norm $\|\cdot\|_{\mA}$ is the $l_1$ norm. When $\mA$ is the set of unit-norm rank-$1$ matrices, the atomic norm is the nuclear norm (sum of singular values).

The atomic norm denoising problem is to use atomic norm regularization to denoise a signal known to be a sparse nonnegative combination of atoms from a set $\mA$. Suppose we have an observation $\y=\x+\n$ and that we know in a priori that $\x$ can be expressed as a linear combination of a small number of atoms from $\mA$. Intuitively, we could search over all the possible short linear combinations from $\mA$ to select the one which minimizes the error $\|\y-\x\|_2^2$. However, this problem is NP-hard. We can estimate the signal through the following convex optimization \cite{bhaskar2011atomic}
\begin{align}\label{eqn: atomic norm denoising algo dfn}
\underset{\x}{\argmin} \frac{1}{2}\|\y-\x\|_2^2+\tau\|\x\|_{\mA},
\end{align}
where $\tau$ is a regularization parameter.

\subsection{Atomic norm of the channel matrix}
In this part, we describe the atomic norm of the channel matrix and review our atomic norm denoising-based channel estimation scheme proposed in \cite{peng15atomic}. Suppose $\{\u_1 \u_2 \dots \u_P\}$ is a set of $K$-dimensional vectors with unit-norm, i.e. $\|\u_p\|_2=1$. We define an atom as
\begin{align*}
A(\theta,\u)=\frac{1}{\sqrt{M}}\a(\theta)\u^*,
\end{align*}
where $\a(\theta)$ follows \eqref{eqn: steering vector} and the associated atomic set is given by
\begin{align}\label{eqn: atomic set of the channel}
\mA=\{A(\theta,\u)|\theta\in[-\pi/2, \pi/2], \|\u\|_2=1\}.
\end{align}
Then, the channel matrix $\H$ can be represented as a non-negative combination of $P$ elements from the atomic set
\begin{align*}
\H=\frac{1}{\sqrt{P}}\A\G=\sum_{p=1}^{P}c_pA(\theta_p, \u_p),
\end{align*}
where $\A$ contains $P$ steering vectors and $c_p=\sqrt{\frac{M}{P}}\|\G_{(p,:)}\|_2$. The atomic norm of an arbitrary matrix $\V$ with respect to $\mA$ can be expressed as
\begin{align*}
\|\V\|_{\mA}&=\inf \{t>0: \V\in t\text{conv}(\mA)\}\\
                   &=\inf \{\sum_{k}c_k | \V=\sum_{k} c_kA(\theta_k, \u_k), c_k\geq 0\},
\end{align*}
where $\text{conv}(\mA)$ is the convex hull of $\mA$. Based on this definition, the atomic norm of the channel matrix can be written as
\begin{align*}
\|\H\|_{\mA}\leq \sum_{p=1}^{P}c_p = \sum_{p=1}^{P}\sqrt{\frac{M}{P}}\|\G_{(p,:)}\|_2.
\end{align*}
For the rest of this paper, the atomic norm $\|\cdot\|_{\mA}$ refers to the one associated with the atomic set \eqref{eqn: atomic set of the channel}.

Next, we review the atomic norm denoising-based channel estimation scheme. For an arbitrary row-wise orthonormal matrix $\PPhi\in\C^{K\times L}$ ($L\geq K$), i.e. $\PPhi\PPhi^*=\I$, our training matrix $\X\in\C^{K\times L}$ is given by
\begin{align}\label{eqn: pilot matrix}
\X=\sqrt{\frac{\totP}{K}}\PPhi.
\end{align}
Multiplying \eqref{eqn: L symbols chn model} by $\PPhi^*$, we obtain
\begin{align}\label{eqn: chn to AD formulation}
\Z\triangleq\frac{\Y\PPhi^*}{\sqrt{\totP/K}}=\H+\N,
\end{align}
where $\N\triangleq\frac{1}{\sqrt{\totP/K}}\N_0\PPhi^*$ has i.i.d. $\mathcal{CN}(0, \frac{K\sigma^2}{\totP})$ entries. Here, the channel estimation is reduced to the estimation of a simple model $\H$ from its noisy observations $\Z$, which turns out to be an atomic norm denoising problem \eqref{eqn: atomic norm denoising algo dfn}. The estimated channel matrix $\Had$ (AD is short for atomic norm denoising) can be obtained by the following atomic norm regularized algorithm:
\begin{align}\label{eqn: ad regularized algorithm}
\Had=\underset{\H}{\argmin}\frac{1}{2}\|\H-\Z\|_F^2+\tau\|\H\|_{\mA},
\end{align}
where $\tau$ denotes a regularization parameter, the Frobenius norm $\|\H-\Z\|_F^2$ is employed to promote the structure of the additive noise and the penalization on the atomic norm is due to the fact that the channel can be expressed as a non-negative combination of a small number of elements from the atomic set (i.e., $P$ is small.). In \cite{li2014off}, it is shown that \eqref{eqn: ad regularized algorithm} can be efficiently implemented via the alternating direction method of multipliers (ADMM). The channel estimation error has been proved in \cite[Proposition 1]{peng15atomic}.

\section{Channel Estimation with Faulty Antennas}\label{sec: chn est with defective antennas}
In this section, we consider massive MIMO systems with failure of arbitrary antennas at the BS. The uncertainty on the indices of faulty antennas and the associated additive corruption noise can affect the channel estimation accuracy, and thus degrade the data transmission performance. We first analyze the signal detection MSE with a linear receiver (MRC, ZF) on the uplink in the large system limit, and show that the effect of faulty antennas does not vanish even with an infinite number of BS antennas. We then prove that a modified receiver can always achieve better signal detection performance provided that the indices of faulty antennas are available. This motivates the study of identifying faulty antennas before the data transmission.

In the second part, we propose an atomic norm-based method that can estimate the channel and detect faulty antennas simultaneously in the training phase. By treating the channel matrix as a low rank matrix, a stable principle component pursuit (PCP)-based approach can also be applied. We provide a comparison between the proposed method and the stable PCP-based one.

Lastly, we present a fast algorithm and prove that it is a good approximation of the atomic norm-based method.

\subsection{Asymptotic Analysis}\label{subsec: asymptotic analysis}
In the UL data transmission, the received signal is
\begin{align}
\y=\sqrt{\rho}\H\x+\w+\n,
\end{align}
where $\rho$ is the signal transmit power, the transmitted signal from $K$ users $\x\in\C^{K}$ is a random vector with zero mean and covariance $\I_K$, the sparse vector $\w\in\C^{M}$ denotes the corruption error due to faulty antennas at the BS, and $\n\in\C^{M}$ is the independent additive noise with $\n\sim\mathcal{CN}(0,\I_M)$.

We employ MRC (or ZF) receiver for the signal detection at the BS, and assume perfect CSI $\H$ for the asymptotic analysis. The mean squared error of signal detection $\E\{\|\x-\xh\|_2^2\}$ is used as the performance measure, where the expectation is with respect to $\x$, $\n$, and the channel matrix $\H$. We assume that $\x$, $\n$, and the channel matrix $\H$ are independent from each other. We consider the large system limit, where $S$ and $M$ grow infinitely large while keeping the ratio $\gamma=S/M$ fixed. In other words, the number of faulty antennas increases proportionally with the number of total antennas. Here, $M\rightarrow\infty$ means $M\rightarrow\infty$, $S\rightarrow\infty$, while $\gamma$, $K$ and $P$ are fixed.

We first present the asymptotic analysis in detail on MRC receiver. Then, the same technique is employed to analyze the signal detection performance of ZF receiver.
\subsubsection{Asymptotic analysis on (modified) MRC receiver}
Two important preliminaries:
\begin{lemma}\cite[Lemma 3.2]{tse2000linear}\label{lem: expect to trace}
Let $\s=[s_1\dots s_N]^T$ where $s_i$'s are i.i.d. zero mean, unit variance random variables with finite fourth moment. Let $\A$ be a deterministic symmetric positive definite matrix. Then
\begin{align}\label{eqn: expect to trace}
\E\{\s^*\A\s\}=\tr\A
\end{align}
\end{lemma}

Based on the finite scattering channel model with $\A=[\a(\theta_1), \a(\theta_2), \cdots, \a(\theta_P)]\in\C^{M\times P}$ containing $P$ steering vectors, we have \cite[Equation (27-28)]{NHQ13Massive}, for $p\neq q$,
\begin{align*}
\frac{1}{M}\a(\theta_p)^*\a(\theta_q)&=\frac{1}{M}\sum_{m=0}^{M-1}e^{j2\pi\frac{D}{\lambda}(\sin\theta_p-\sin\theta_q)m}\\
                                     &=\frac{1}{M}\frac{1-e^{j2\pi\frac{D}{\lambda}(\sin\theta_p-\sin\theta_q)M}}{1-e^{j2\pi\frac{D}{\lambda}(\sin\theta_p-\sin\theta_q)}}\\
                                     &\overset{M\rightarrow\infty}{\rightarrow}0,
\end{align*}
and
\begin{align*}
\frac{1}{M}\a(\theta_p)^*\a(\theta_p)=1.
\end{align*}
In other words,
\begin{align}\label{eqn: AA to identity}
\frac{1}{M}\A^*\A\overset{M\rightarrow\infty}{\rightarrow}\I_P
\end{align}

By MRC, the estimation on the transmitted signal $\x$ is
\begin{align}\label{eqn: MRC receiver}
\xh=\frac{\H^*\y}{\sqrt{\rho}M}=\frac{\H^*\H}{M}\x+\frac{\H^*\w}{\srho M}+\frac{\H^*\n}{\srho M}
\end{align}

The MSE can be decomposed as \eqref{eqn: mse decomp 1} at the top of the next page.
\begin{figure*}[!t]
\normalsize
\setcounter{MYtempeqncnt}{\value{equation}}
\setcounter{equation}{17}
\begin{align}
\E\{\|\x-\xh\|_2^2\}&=\E\{(\x-\xh)^*(\x-\xh)\}\nonumber\\
                    &=\E\left\{(\x-\frac{\H^*\H}{M}\x-\frac{\H^*\w}{\srho M}-\frac{\H^*\n}{\srho M})^*(\x-\frac{\H^*\H}{M}\x-\frac{\H^*\w}{\srho M}-\frac{\H^*\n}{\srho M})\right\}\nonumber\\
                    &=\E\left\{\|\x\|_2^2-2\frac{\x^*\H^*\H\x}{M}-2\frac{\x^*\H^*\w}{\srho M}-2\frac{\x^*\H^*\n}{\srho M}+\frac{\x^*\H^*\H\H^*\H\x}{M^2}+2\frac{\x^*\H^*\H\H^*\w}{\srho M^2}\right.\nonumber\\
                    &\left.\quad\quad\quad+2\frac{\x^*\H^*\H\H^*\n}{\srho M^2}+\frac{\w^*\H\H^*\w}{\rho M^2}+2\frac{\w^*\H\H^*\n}{\rho M^2}+\frac{\n^*\H\H^*\n}{\rho M^2}\right\}\label{eqn: mse decomp 1}
\end{align}
\setcounter{equation}{\value{MYtempeqncnt}}
\addtocounter{equation}{1}
\hrulefill
\vspace*{4pt}
\end{figure*}

Due to independence and \eqref{eqn: AA to identity}
\begin{align*}
\E\left\{\frac{\x^*\H^*\w}{M}\right\}&=0,\quad \E\left\{\frac{\x^*\H^*\n}{M}\right\}=0,\\
\E\left\{\frac{\x^*\H^*\H\H^*\w}{M^2}\right\}&=0,\quad \E\left\{\frac{\x^*\H^*\H\H^*\n}{M^2}\right\}=0,\\
\E\left\{\frac{\w^*\H\H^*\n}{M^2}\right\}&=0.
\end{align*}

We have
\begin{align}
\E\left\{\frac{\n^*\H\H^*\n}{M^2}\right\}&=\frac{\E\{\tr(\H\H^*)\}}{M^2}\nonumber\\
                                         &=\frac{\E\{\tr(\A\G\G^*\A^*)\}}{PM^2}\nonumber\\
                                         &=\frac{\tr(\A\E\{\G\G^*\}\A^*)}{PM^2}\nonumber\\
                                         &=\frac{\tr(\A\A^*)K}{PM^2}\nonumber\\
                                         &=\frac{MPK}{PM^2}\overset{M\rightarrow\infty}{\rightarrow}0,
\end{align}
where the first equality is due to Lemma \ref{lem: expect to trace}.

\begin{align}
\E\{\|\x\|_2^2\}&=K\\
\E\left\{\frac{\x^*\H^*\H\x}{M}\right\}&=\frac{1}{P}\E\left\{\frac{\x^*\G^*\A^*\A\G\x}{M}\right\}\nonumber\\
                                       &=\frac{1}{P}\tr(\frac{\E\{\G^*\A^*\A\G\}}{M})\nonumber\\
                                       &\overset{M\rightarrow\infty}{\rightarrow}\frac{1}{P}\tr(\E\{\G^*\G\})=K,
\end{align}
where the last two steps are due to Lemma \ref{lem: expect to trace} and \eqref{eqn: AA to identity} respectively.

Similarly,
\begin{align}\label{eqn: equation related to P}
\E\left\{\frac{\x^*\H^*\H\H^*\H\x}{M^2}\right\}&=\frac{\E\{\tr(\H^*\H\H^*\H)\}}{M^2}\nonumber\\
                                               &=\frac{\E\{\tr(\G^*\A^*\A\G\G^*\A^*\A\G)\}}{P^2M^2}\nonumber\\
                                               &\overset{M\rightarrow\infty}{\rightarrow}\frac{\E\{\tr(\G^*\G\G^*\G)\}}{P^2}\nonumber\\
                                               &=\frac{K^2}{P}+\frac{K}{P}+K,
\end{align}

\begin{align}
\E\left\{\frac{\w^*\H\H^*\w}{M^2}\right\}&=\frac{\E\{\w^*\A\G\G^*\A^*\w\}}{PM^2}\nonumber\\
                                         &=\frac{\w^*\A\E\{\G\G^*\}\A^*\w}{PM^2}\nonumber\\
                                         &=\frac{K}{PM^2}\|\A^*\w\|_2^2.
\end{align}

Combine the above equations, we have
\begin{align}\label{eqn: mse perfect antenna}
\E\{\|\x-\xh\|_2^2\}\minfty\frac{K^2}{P}+\frac{K}{P}+\frac{K}{\rho PM^2}\|\A^*\w\|_2^2.
\end{align}
For any $\a(\theta_i)$, $i\in[P]$, we have $\a(\theta_i)^*\w\leq\gamma M\|\w\|_{\infty}$. Hence,
\begin{align}\label{eqn: difference of the MSEs}
0\leq\frac{K}{\rho PM^2}\|\A^*\w\|_2^2\leq\frac{K\gamma^2}{\rho}\|\w\|_{\infty}^{2}.
\end{align}

Next, we analyze the asymptotic MSE when the indices of faulty antennas are known. We denote the support of $\w$ by $\Omega\in\{1,\dots,M\}$, i.e., $|\Omega|=S=\gamma M$. Suppose $\Omega$ is known in a priori at the BS, we shall prove that this knowledge can be utilized to obtained better signal detection performance in terms of asymptotic MSE. Let $\Ab$ denote the matrix obtained by removing the rows of $\A$ indexed by $\Omega$, $\Ab\in\C^{(1-\gamma)M\times P}$, and $\nb\in\C^{(1-\gamma)M}$ represent the i.i.d. additive noise vector with zero mean and unit variance.

Since the indices of faulty antennas $\Omega$ are known, we can omit the readings received from those antennas at the BS. In this way, the impact of $\w$ is removed and the received signal is
\begin{align}\label{eqn: received sg after remv}
\yb=\srho\Hb\x+\nb,
\end{align}
where $\Hb:=\frac{1}{\sqrt{P}}\Ab\G$.

By MRC, the transmitted signal can be recovered by
\begin{align*}
\xb=\frac{\Hb^*\yb}{(1-\gamma)\srho M}=\frac{\Hb^*\Hb\x}{(1-\gamma)M}+\frac{\Hb^*\nb}{(1-\gamma)M\srho}
\end{align*}

For $\Ab=[\ab(\theta_1), \ab(\theta_2), \cdots, \ab(\theta_P)]$ with $\ab(\theta_i)$'s being the vectors obtained by removing the $\Omega$-th entries of $\a(\theta_i)$, we have, for $p\neq q$,
\begin{align*}
\frac{1}{(1-\gamma)M}\ab(\theta_p)^*\ab(\theta_q)&=\frac{1}{(1-\gamma)M}\sum_{m\in[M]\setminus\Omega}e^{j2\pi\frac{D}{\lambda}(\sin\theta_p-\sin\theta_q)m}\\
                                     &\overset{M\rightarrow\infty}{\rightarrow}0,
\end{align*}
and
\begin{align*}
\frac{1}{(1-\gamma)M}\ab(\theta_p)^*\ab(\theta_p)=1.
\end{align*}
Hence,
\begin{align}\label{eqn: AA to identity 2}
\frac{1}{(1-\gamma)M}\Ab^*\Ab\overset{M\rightarrow\infty}{\rightarrow}\I_P
\end{align}

Similarly, the asymptotic MSE can be obtained by applying Lemma \ref{lem: expect to trace} and \eqref{eqn: AA to identity 2}
\begin{align}\label{eqn: mse defective antenna}
&\E\{\|\x-\xb\|_2^2\}\nonumber\\
&=\E\{(\x-\xb)^*(\x-\xb)\}\nonumber\\
&=\E\left\{\|\x\|_2^2-2\frac{\x^*\Hb^*\Hb\x}{(1-\gamma)M}-2\frac{\x^*\H^*\nb}{(1-\gamma)M\srho}+\frac{\x^*\Hb^*\Hb\Hb^*\Hb\x}{(1-\gamma)^2M^2}\right.\nonumber\\
&\left.+2\frac{\x^*\Hb^*\Hb\Hb^*\nb}{(1-\gamma)^2M^2\srho}+\frac{\nb^*\Hb\Hb^*\nb}{(1-\gamma)^2M^2\srho}\right\}\nonumber\\
&\minfty\frac{K^2}{P}+\frac{K}{P}
\end{align}

Hence, in the large system limit, the MSE obtained by appealing to the knowledge on the indices of faulty antennas is always smaller than the one given by the conventional method (compare \eqref{eqn: mse perfect antenna} and \eqref{eqn: mse defective antenna}). Some remarks are in order.

 \begin{enumerate}
   \item The difference of the two MSEs ranges from $0$ to $\frac{K\gamma^2}{\rho}\|\w\|_{\infty}^{2}$ \eqref{eqn: difference of the MSEs}. The result is consistent with the intuitive understanding: when $\gamma=0$, i.e. all antennas are fault-free, the two MSEs are identical; in the worst case, the difference $\frac{K\gamma^2}{\rho}\|\w\|_{\infty}^{2}$ is proportional to the number of users $K$ and inversely proportional to the signal transmit power $\rho$.
   \item In the large system limit with fault-free antennas, MRC has been shown to mitigate the effect of uncorrelated noise and small-scale fading \cite{marzetta2010noncooperative}. When considering faulty-antennas, our analysis indicates that the impact of faulty-antennas does not vanish with unlimited number of antennas. However, with the knowledge on the indices of faulty antennas, this effect can be eliminated by a simple modified MRC receiver.
   \item In this paper, our analysis does not rely on any assumption on the statistical model of the distortion noise caused by faulty antennas. If the statistical information on either the positions of faulty antennas or the magnitude of the distortion caused by faulty antennas is available/etimatable, other receivers (e.g. MMSE) could be investigated.
   \item When the channel matrix $\H$ is i.i.d. Gaussian, similar result can be obtained by following the derivations above.
 \end{enumerate}
\subsubsection{Asymptotic analysis on (modified) ZF receiver}\label{subsec: improved receiver2}
With a similar technique, we shall show that a modified ZF receiver with the knowledge on the indices of faulty antennas can provide better signal detection performance in the large system limit. The ZF receiver can suppress interuser interference and perform well at high SNR. For the ZF receiver, we assume that $P\geq K$ \footnote{This is to ensure that $\G^*\G$ is invertible. The same assumption has been used in the literature, e.g. \cite{NHQ13Massive}.}. Here, we use $\E\{\|\x-\xh\|_2\}$ as the performance metric for convenience. Without a priori knowledge on the indices of faulty antennas, the signal recovered by ZF receiver can be written as
\begin{align*}
\xh=\frac{\H^{\dagger}\y}{\sqrt{\rho}}=\x+\frac{\H^{\dagger}\w}{\sqrt{\rho}}+\frac{\H^{\dagger}\n}{\sqrt{\rho}},
\end{align*}
where $\H^{\dagger}\triangleq(\H^*\H)^{-1}\H^*$. Denoting $\G^{\dagger}\triangleq(\G^*\G)^{-1}\G^*$, the asymptotic error is given by
\begin{align*}
\E\{\|\x-\xh\|_2\}&\leq\frac{1}{\sqrt{\rho}}\E\{\|\H^{\dagger}\w\|_2\}+\E\{\|\H^{\dagger}\n\|_2\}\\
                    &=\frac{1}{\sqrt{\rho}}\E\{\left\|(\frac{\G^*\A^*\A\G}{P})^{-1}\frac{\G^*\A^*}{\sqrt{P}}\w\right\|_2\}\\
                    &=\frac{\sqrt{P}}{\sqrt{\rho} M}\E\{\left\|(\frac{\G^*\A^*\A\G}{M})^{-1}\G^*\A^*\w\right\|_2\}\\
                    &\minfty\frac{\sqrt{P}}{\sqrt{\rho} M}\E\{\left\|(\G^*\G)^{-1}\G^*\A^*\w\right\|_2\}\\
                    &\leq\frac{\sqrt{P}}{\sqrt{\rho} M}\E\{\left\|\G^{\dagger}\right\|_2\}\|\A^*\w\|_2\\
                    &\overset{(a)}{\leq}\frac{\sqrt{P}}{\sqrt{\rho} M(\sqrt{P}-\sqrt{K})}\|\A^*\w\|_2\\
                    &\overset{(b)}{\leq}\frac{\gamma P \|\w\|_\infty}{\sqrt{\rho}(\sqrt{P}-\sqrt{K})},
\end{align*}
where (a) is based on Gordon's theorem for Gaussian matrices (\cite[Theorem 5.32]{vershynin2010introduction}) and the fact that $s_{min}(\G)=1/\|\G^{\dagger}\|$ with $s_{min}(\G)$ denoting the smallest singular value of $\G$; (b) is due to a similar technique to obtain \eqref{eqn: difference of the MSEs}. Like the modified MRC receiver, we can omit the readings from the faulty antennas if their locations are available, and apply ZF receiver to the signal \eqref{eqn: received sg after remv}. We have
\begin{align*}
\xb=\frac{\Hb^{\dagger}\yb}{\sqrt{\rho}}=\x+\frac{\Hb^{\dagger}\nb}{\sqrt{\rho}}.
\end{align*}
The asymptotic error is $\E\{\|\x-\xb\|_2\}\minfty0$, which demonstrates the improved signal detection performance of the modified ZF receiver.

\begin{remark}
\emph{Consider the case when the number of dominant AoAs $P$ also grows infinitely large and assume that $M$ grows at a greater rate than $P$ (as in \cite[Remark 1]{NHQ13Massive}). Here, $M,P\rightarrow\infty$ means $M\rightarrow\infty$, $S\rightarrow\infty$, $P\rightarrow\infty$, while $\gamma$ and $K$ are fixed. We have
\begin{align*}
\frac{1}{M}\A^*\A &\overset{M,P\rightarrow\infty}{\rightarrow}\I_P,\\
\frac{1}{P}\G^*\G &\overset{M,P\rightarrow\infty}{\rightarrow}\I_K.
\end{align*}
For the MRC receiver without the knowledge on the indices of the faulty antennas, the affected term is \eqref{eqn: equation related to P}, which is now given by
\begin{align}
\E\left\{\frac{\x^*\H^*\H\H^*\H\x}{M^2}\right\}&\overset{M,P\rightarrow\infty}{\rightarrow}K.
\end{align}}

\emph{Therefore, we have $\E\{\|\x-\xh\|_2^2\}\overset{M,P\rightarrow\infty}{\rightarrow}\frac{K}{\rho PM^2}\|\A^*\w\|_2^2$. It is noted that this term is bounded by \eqref{eqn: difference of the MSEs}. If the indices of the faulty antennas are available, it can be proven that $\E\{\|\x-\xb\|_2^2\}\overset{M,P\rightarrow\infty}{\rightarrow}0$ by following the same steps.}

\emph{For the ZF receiver, it can be shown that the ZF receiver is equivalent to the MRC receiver when $M,P\rightarrow\infty$ and $M$ growing at a greater rate than $P$. We have
\begin{align*}
\xh&=\frac{\H^{\dagger}\y}{\sqrt{\rho}}\\
     &=\frac{(\H^*\H)^{-1}\H^*\y}{\sqrt{\rho}}\\
     &=\frac{1}{M}(\frac{\G^*\A^*\A\G}{MP})^{-1}\frac{\H^*\y}{\sqrt{\rho}}\\
     &\overset{M,P\rightarrow\infty}{\rightarrow}\frac{\H^*\y}{\sqrt{\rho}M},
\end{align*}
which is the same as the estimation given by the MRC receiver \eqref{eqn: MRC receiver}. Hence, the asymptotic signal detection MSE follows the results for the MRC receiver.}
\end{remark}

\subsection{The proposed method}
Since the magnitude of the sparse distortion $\w$ may change arbitrarily for different symbol intervals, estimation of the exact vector $\w$ that affects data transmission is impossible. From the asymptotic analysis in the previous subsection, we may estimate the channel matrix and detect the indices of faulty antennas simultaneously in the training phase and apply the modified MRC/ZF receiver to guarantee signal detection performance. We therefore propose the following method.

We assume that the training matrix $\X$ is constructed as \eqref{eqn: pilot matrix}. Multiplying \eqref{eqn: L symbols chn model distortion} by $\PPhi^*$, we obtain
\begin{align}\label{eqn: matrix decomp}
\Z\triangleq\frac{\Y\PPhi^*}{\sqrt{\totP/K}}=\H+\W+\N,
\end{align}
where $\N\triangleq\frac{1}{\sqrt{\totP/K}}\N_0\PPhi^*$ has i.i.d. $\mathcal{CN}(0, \frac{K\sigma^2}{\totP})$ entries and $\W\triangleq\frac{1}{\sqrt{\totP/K}}\W_0\PPhi^*\in\C^{M\times K}$ is still a matrix with $S$ nonzero rows. Also, the indices of nonzero rows of $\W$ are the same as those of $\W_0$. Now, our problem is to decompose the given matrix $\Z$ into a sum of three component matrices, each of which is `simple' in a sense described by the corresponding norm. To recover $\H$ and $\W$, we propose to solve the following \emph{extended} atomic norm minimization algorithm:
\begin{align}\label{eqn: extended AD}
\underset{\H,\W}{\argmin}\ \frac{1}{2}\|\H+\W-\Z\|_{F}^2+\tau_1\|\H\|_{\mA}+\tau_2\|\W\|_{\mnorm}.
\end{align}
Here, the function $\|\cdot\|_{\mnorm}$ representing the sum of the Euclidean norms of the rows of the matrix is applied since $\W$ is a matrix with a small number of nonzero rows. Although the dimensions of $\W_0$ and $\W$ are different, recovering $\W$ is good enough since its indices of nonzero rows are the same as those of $\W_0$. To implement the proposed algorithm, we reformulate it into a matrix decomposition problem
\begin{align}
\underset{\H,\W,\N}{\argmin}&\ \frac{1}{2}\|\N\|_{F}^2+\tau_1\|\H\|_{\mA}+\tau_2\|\W\|_{\mnorm}\\
\text{s.t.}&\quad \H+\W+\N=\Z\nonumber,
\end{align}
which can be solved based on exchange and ADMM \cite{parikh2013proximal}. Detail steps are included in Algorithm \ref{alg: implementation}. In every iteration, each of the matrices $\H,\W,\N$ is updated by a convex optimization algorithm. As shown in Algorithm \ref{alg: implementation}, the updates on $\W$ and $\N$ have simple closed form solutions. While for the update on $\H$,
\begin{align}\label{eqn: prox algo atomic norm}
\H=\underset{\H}{\argmin}\ \tau_1\|\H\|_{\mA}+\frac{1}{2}\|\H-\V_1^k\|_{F}^2
\end{align}
is exactly the atomic norm regularization algorithm in \eqref{eqn: ad regularized algorithm}, which can be efficiently implemented by the ADMM as described in \cite{li2014off}.

\begin{remark}
\emph{
In the case of $K=1$, the problem \eqref{eqn: matrix decomp} is reduced to the line spectrum estimation of a vector signal that is corrupted by sparse noise and dense noise. Currently, the theoretical guarantee has been proved only when the dense noise is absent \cite{chiconvex,fernandez2016demixing}. When both the sparse noise and the dense noise exist, a vector version of the proposed extended atomic norm minimization algorithm is adapted to recover the signal \cite{fernandez2016demixing}. In addition, the vector version has been applied in the delay-doppler estimation for OFDM passive radar \cite{zheng2016super}.
}
\end{remark}

\begin{algorithm}
\caption{Implementation of the extended atomic norm minimization algorithm}\label{alg: implementation}
\begin{algorithmic}
\Require $\Z$, $\tau_1$, $\tau_2$
\Ensure $\H^{(0)}$, $\W^{(0)}$, $\N^{(0)}$, $\U^{(0)}$
\For{$k=1,2,\dots$}
\State
\begin{align*}
\Xb^k&=\frac{1}{3}(\H^k+\W^k+\N^k)\\
\V_1^k&=\H^k-\Xb^k+(1/3)\Z-\U^k\\
\V_2^k&=\W^k-\Xb^k+(1/3)\Z-\U^k\\
\V_3^k&=\N^k-\Xb^k+(1/3)\Z-\U^k\\
\H^{k+1}&=\underset{\H}{\argmin}\ \tau_1\|\H\|_{\mA}+\frac{1}{2}\|\H-\V_1^k\|_{F}^2\\
\W^{k+1}&=\underset{\W}{\argmin}\ \tau_2\|\W\|_{\mnorm}+\frac{1}{2}\|\W-\V_2^k\|_{F}^2\\
\N^{k+1}&=\underset{\N}{\argmin}\ \frac{1}{2}\|\N\|_{F}^2+\frac{1}{2}\|\N-\V_3^k\|_{F}^2\\
\Xb^{k+1}&=\frac{1}{3}(\H^{k+1}+\W^{k+1}+\N^{k+1})\\
\U^{k+1}&=\U^k+\Xb^{k+1}-(1/3)\Z
\end{align*}
\State Solve for $\H^{k+1}$ by another ADMM algorithm as in \cite{li2014off}
\State Solve for $\W^{k+1}$ and $\N^{k+1}$
\begin{align*}
\W_{(m,:)}^{k+1}&=(1-\frac{\tau_2}{\|(\V_2^k)_{(m,:)}\|_2})_{+}(\V_2^k)_{(m,:)}\\
\N^{k+1}&=\frac{1}{2}\V_3^k
\end{align*}
\State \ \Comment{$m=1,\dots,M$}
\State \ \Comment{$(x)_{+}=x$ if $x\geq 0$ and $(x)_{+}=0$ otherwise}
\EndFor
\State \Return $\H$, $\W$
\end{algorithmic}
\end{algorithm}

\subsection{The stable PCP-based approach}
In the proposed method, we can simultaneously estimate the channel and detect faulty antennas by mainly exploring the knowledge that the intrinsic information of the channel matrix is small, that is the channel matrix is a linear combination of a small number of elements from the atomic set \eqref{eqn: atomic set of the channel}. We note that similar idea of exploring the intrinsic information for channel estimation has been considered based on compressed sensing. In \cite{nguyen2013compressive,rao2014distributed,migliore2016minimum}, where no faulty antenna is considered, channel estimation is transformed into a low-rank matrix recovery problem by noticing the fact that the channel matrix is low rank for finite scattering channel models. Comparing with our proposed method, it naturally leads to another algorithm for channel estimation and faulty antenna detection. By relying on the knowledge that the channel matrix $\H$ is low rank, we can solve the matrix decomposition problem \eqref{eqn: matrix decomp} by the following algorithm
\begin{align}\label{eqn: stPCP}
\underset{\H,\W,\N}{\argmin}&\ \frac{1}{2}\|\N\|_{F}^2+\lambda_1\|\H\|_{*}+\lambda_2\|\W\|_{\mnorm}\\
\text{s.t.}&\quad \H+\W+\N=\Z\nonumber,
\end{align}
where $\|\cdot\|_{*}$ denotes the nuclear norm. In other words, the algorithm is to recover a low rank matrix which is corrupted by a row-sparse matrix and additive noise. We note that the above algorithm is quite similar to the stable principle component pursuit (PCP) algorithm \cite{zhou2010stable}. The only difference is that in \cite{zhou2010stable} $\W$ is a sparse matrix, whereas here it is a row-sparse matrix. This stable PCP-based algorithm \eqref{eqn: stPCP} can be efficiently implemented via ADMM\footnote{In the literature, the stable PCP can also be solved by standard SDP and fast proximal gradient algorithm. However, the state-of-art solver is by ADMM \cite{zhou2010stable,Aybat15alternating}.}. The update steps are in general the same as those of our proposed method (see Algorithm \ref{alg: implementation}). The difference is on the update of $\H$, where the following algorithm is implemented,
\begin{align}\label{eqn: prox algo nuclear norm}
\H=\underset{\H}{\argmin}\ \lambda_1\|\H\|_{*}+\frac{1}{2}\|\H-\V_1^k\|_{F}^2.
\end{align}

\subsection{A fast algorithm}
Comparing with LS-based channel estimation methods, the proposed atomic norm-based method has higher computational complexity. The computational complexity for LS and LS-SLS methods are $\bigo(MLK+K^3)$ and $\bigo(M^2L+MLK)$ respectively. As shown in Algorithm \ref{alg: implementation}, the main step of our proposed algorithm is to update the estimated channel matrix $\H$ \eqref{eqn: prox algo atomic norm}, while the rest updates have closed form solutions. The dominate operation in each iteration for solving \eqref{eqn: prox algo atomic norm} is the eigenvalue decomposition of a $(M+K)\times (M+K)$ matrix, hence the complexity is $\bigo((M+K)^3)$. Similarly, for the stable PCP-based approach \eqref{eqn: stPCP}, the main step is also to update the estimated channel matrix by solving \eqref{eqn: prox algo nuclear norm}. In this algorithm, the dominate operation in each iteration is the singular value decomposition (SVD) for a $M\times K$ matrix, which has computational complexity $\bigo(M^2K+K^3)$. We note that the computational complexity of the stable PCP-based approach is slightly better than our proposed method, however, its performance is degraded. In this part, we present a fast algorithm for the proposed atomic norm-based approach. We prove that the fast algorithm performs as well as the atomic norm-based approach, and has a much lower computational complexity than both the atomic norm and stable PCP-based approaches.

Recall that the steering vector associated with each angle-of-arrival (AOA) is given by
\begin{align*}
\a(\theta) \triangleq [1 \quad \mathrm{e}^{-j2\pi\frac{D}{\lambda}\sin(\theta)} \dots \quad \mathrm{e}^{-j2\pi\frac{(M-1)D}{\lambda}\sin(\theta)}]^T,
\end{align*}
where $\theta\in[-\pi/2, \pi/2]$ and $\frac{D}{\lambda}<1$. For each $\theta\in[-\pi/2, \pi/2]$, we can find a corresponding $f\in[0,1)$ such that the steering vector can be written as
\begin{align*}
\a(f) \triangleq [1 \quad \mathrm{e}^{j2\pi f} \dots \quad \mathrm{e}^{j2\pi(M-1)f}]^T.
\end{align*}
Accordingly, the atomic set in \eqref{eqn: atomic set of the channel} can be expressed as
\begin{align}\label{eqn: atomic set of the channel with f}
\mA=\{A(f,\u)=\frac{1}{\sqrt{M}}\a(f)\u^*|f\in[0,1), \|\u\|_2=1\}.
\end{align}
To present the fast algorithm, we divide the interval $[0,1)$ into $N$ uniform grids and define a new atomic set
\begin{align}\label{eqn: atomic set of the channel with f and N}
\mA_N=\{A_N(f,\u)|f\in\{0,\frac{1}{N},\dots,\frac{N-1}{N}\}, \|\u\|_2=1\},
\end{align}
where $A_N(f,\u)=\frac{1}{\sqrt{M}}\a(f)\u^*$. Based on the new atomic set, the channel matrix can be written as
\begin{align}
\H=\frac{1}{\sqrt{P}}\overbar{\F}\overbar{\G}=\sum_{i=1}^{N}c_iA_N(f_i,\u_i),
\end{align}
where $\overbar{\F}=[\frac{1}{\sqrt{M}}\a(0), \dots, \frac{1}{\sqrt{M}}\a(\frac{N-1}{N})]\in\C^{M\times N}$ ($M<N$) is the first $M$ rows of a normalized discrete Fourier transform (DFT) matrix, $\overbar{\G}\in\C^{N\times K}$ and $c_i=\sqrt{\frac{M}{P}}\|\overbar{\G}_{(i,:)}\|_2$. If the channel matrix is related to a finite scattering channel model with $P$ paths, the above expression indicates that $\overbar{\G}$ is a matrix with $P$ non-zero rows. The atomic norm of the channel matrix associated with the new atomic set is given by
\begin{align}
\|\H\|_{\mA_N}&=\inf\{\sum_{i}c_i|\H=\sum_{i}c_iA_N(f_i,\u_i)\}\\
              &=\inf\{\sum_{i}\sqrt{\frac{M}{P}}\|\overbar{\G}_{(i,:)}\|_2|\H=\frac{1}{\sqrt{P}}\overbar{\F}\overbar{\G}\}\\
              &=\inf\{\sqrt{\frac{M}{P}}\|\overbar{\G}\|_{2,1}|\H=\frac{1}{\sqrt{P}}\overbar{\F}\overbar{\G}\}.
\end{align}
Now, we can solve the matrix decomposition problem \eqref{eqn: matrix decomp} by the following algorithm
\begin{align}\label{eqn: fsAD}
\underset{\H,\W,\N}{\argmin}&\ \frac{1}{2}\|\N\|_{F}^2+\nu_1\|\H\|_{\mA_N}+\nu_2\|\W\|_{\mnorm}\\
\text{s.t.}&\quad \H+\W+\N=\Z\nonumber.
\end{align}
One important question is that when the frequencies $f\in[0,1)$ of the channel are not on the uniform grids, the new atomic norm $\|\H\|_{\mA_N}$ will deviate from the original one $\|\H\|_{\mA}$ and lead to additional estimation error. Here, we address this question by proving that $\|\H\|_{\mA_N}$ is a good approximation of $\|\H\|_{\mA}$ by setting $N$ a constant factor larger than $MK$. In particular, we have (See Appendix \ref{app: bound on new atomic norm} for the proof)
\begin{align}\label{eqn: bound on new atomic norm}
(1-\frac{2\pi MK}{N})^\frac{1}{2}\|\H\|_{\mA_N}\leq\|\H\|_{\mA}\leq\|\H\|_{\mA_N}.
\end{align}
The above result indicates that $\|\H\|_{\mA_N}$ is a good approximation of $\|\H\|_{\mA}$ when $N=\bigo(MK)$ and the gap between them approaches zero as $N\rightarrow\infty$. We note that the term $(1-\frac{2\pi MK}{N})^\frac{1}{2}$ is not tight, which means $N\geq 2\pi MK$ is not necessarily required in practice.

Next, we proceed to explain the computational efficiency of the new algorithm. Clearly, the algorithm \eqref{eqn: fsAD} can still be implemented by exchange and ADMM, as the one for algorithm \eqref{eqn: extended AD} in Algorithm \ref{alg: implementation}. The key difference is to change the update step of $\H$ to
\begin{align}\label{eqn: prox algo fs atomic norm}
\H=\underset{\H}{\argmin}\ \nu_1\|\H\|_{\mA_N}+\frac{1}{2}\|\H-\V_1^k\|_{F}^2,
\end{align}
which is equivalent to
\begin{align}\label{eqn: prox algo fs atomic norm}
\overbar{\G}&=\underset{\overbar{\G}}{\argmin}\ \nu_1\|\overbar{\G}\|_{2,1}+\frac{1}{2}\|\overbar{\F}\overbar{\G}-\V_1^k\|_{F}^2\\
\H &=\overbar{\F} \overbar{\G},
\end{align}
where the first algorithm is a group lasso for the recovery of a matrix with small number of non-zero rows. It can be solved iteratively with each iteration dominated by the matrix multiplication by $\overbar{\F}$. Hence, the computational complexity of the fast algorithm is $\bigo(MK\log(MK))$ by relying on the efficiency of the FFT.
\begin{figure*}[!t]
\centering
\subfloat[Channel estimation]{\includegraphics[width=3.5in]{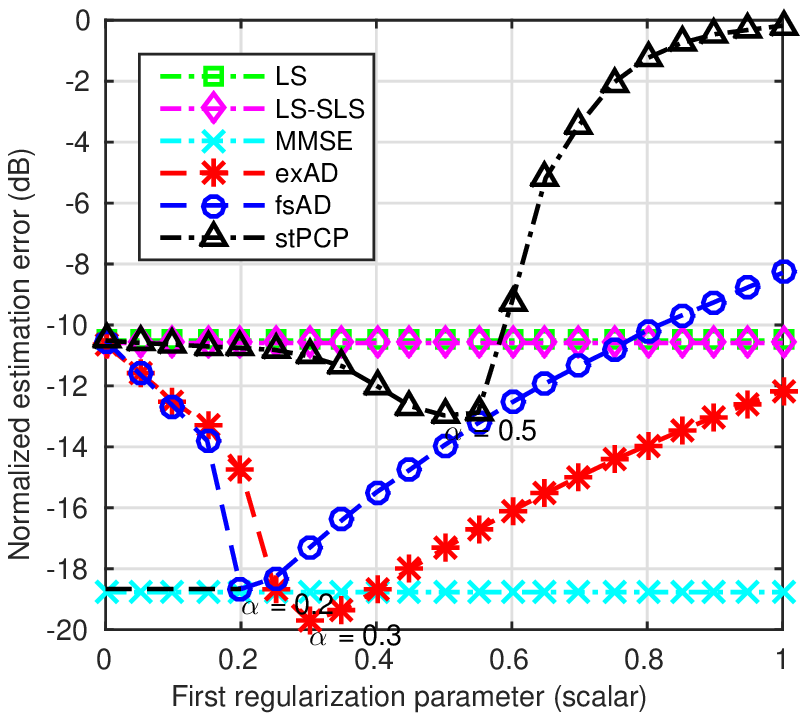}\label{fig: figure 1a}}
\hfil
\subfloat[Faulty antennas detection]{\includegraphics[width=3.5in]{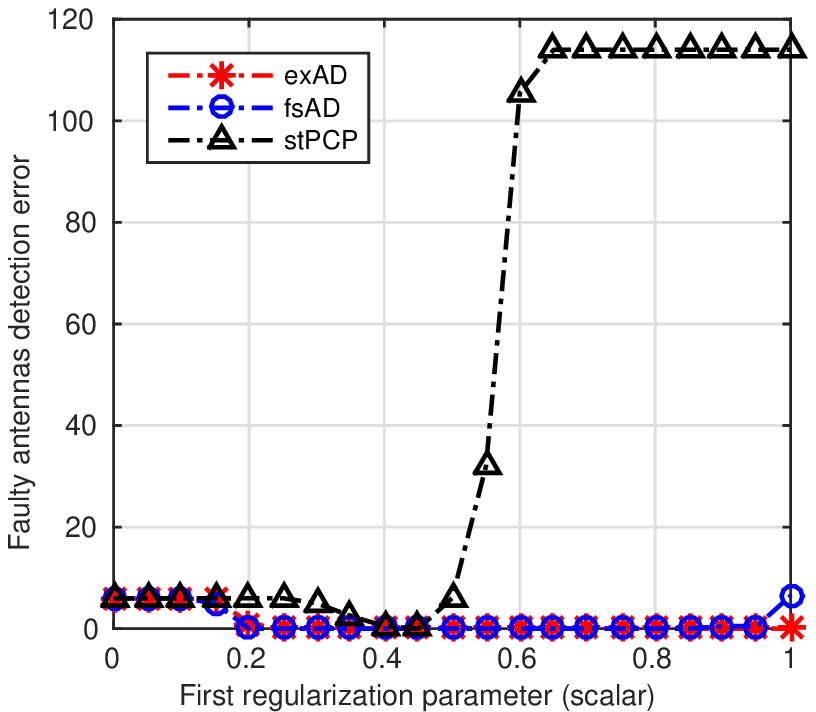}\label{fig: figure 1b}}
\caption{Comparison of the channel estimation methods for different values of $\alpha\in[0:0.05:1]$ with $\{\tau_1,\nu_1,\lambda_1\}=\alpha\|\Z\|_{2}$, $\{\tau_2,\nu_2,\lambda_2\}=0.5\|\Z\|_{\minfnorm}$, $M=120$, $S=\lfloor0.05M\rfloor$, $K=10$, $P=20$ and $L=10$.}
\label{fig: experiment 1}
\end{figure*}

\section{Numerical Results}\label{sec: simulations}
In this section, we conduct numerical experiments on massive MIMO systems with faulty antennas to demonstrate the performance of our proposed extended atomic norm denoising-based method (exAD) \eqref{eqn: extended AD} and the fast atomic norm denoising-based method (fsAD) \eqref{eqn: fsAD}. We also include the simulation results of the LS-based methods, the MMSE method and the stable PCP-based algorithm (stPCP) \eqref{eqn: stPCP} for comparison.

We consider the signal model in \eqref{eqn: L symbols chn model distortion} and the pilot matrix given by \eqref{eqn: pilot matrix}. The additive noise term $\N_0$ is a $M\times L$ matrix with i.i.d. $\mathcal{CN}(0,\sigma^2)$ entries. For the distortion matrix $\W_0$, the indices of the nonzero rows are selected uniformly at random and each nonzero entry is an i.i.d. Bernoulli random variable scaled by $4\sqrt{\rho}$, where $\rho$ represents the power of the transmit symbol. Here, the setting of the distortion matrix is rather arbitrary and aims at generating the synthetic data for the distortion noise caused by faulty antennas. The magnitude $4\sqrt{\rho}$ is selected such that the distortion noise caused at any faulty antenna is proportional to the signal power received at that antenna. However, we remark that our proposed method does not rely on the distribution of nonzero rows or entries of the distortion matrix. The steering vector is set to $D/\lambda=0.3$, AOA $\theta_p=-\pi/2+(p-1)\pi/P$, $p=1,2,\dots,P$, as in \cite{hoydis2013massive}\cite{ngo2011analysis}\cite{nguyen2013compressive}. The channel estimation error (dB) is measured by the following normalized norm
\begin{align}\label{eqn: normailized error}
10\log_{10}(\frac{1}{MK}\|\H-\Hh\|_F^2).
\end{align}

To measure the performances of different algorithms on faulty antenna detection, we define the faulty antenna detection error in the following way. Associated with the distortion matrix $\W_0$, we define a length-$M$ binary vector $\s$ with entries $1$ indicating the locations of faulty antennas and $0$ for fault-free antennas. For example, if only the second row of $\W_0$ is nonzero, then we have $\s={0,\ 1,\ 0,\dots,\ 0}$. Similarly, we can obtain an estimated binary vector $\sh$ based on the estimated distortion matrix $\Wh$. The faulty antenna detection error is defined as the hamming distance between $\s$ and $\sh$, i.e., $\text{hamming}(\s,\sh)$, which accounts for both miss detection and false alarm. Hence, the value of the faulty antenna detection error can range from $0$ to $M$.

\begin{figure*}[!t]
\centering
\subfloat[Channel estimation]{\includegraphics[width=3.5in]{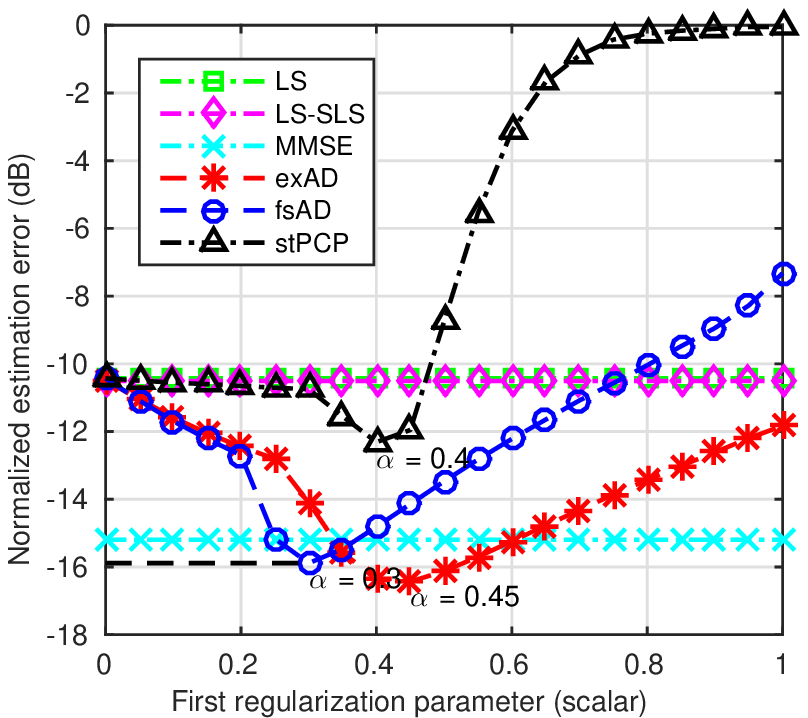}\label{fig: figure 2a}}
\hfil
\subfloat[Faulty antennas detection]{\includegraphics[width=3.5in]{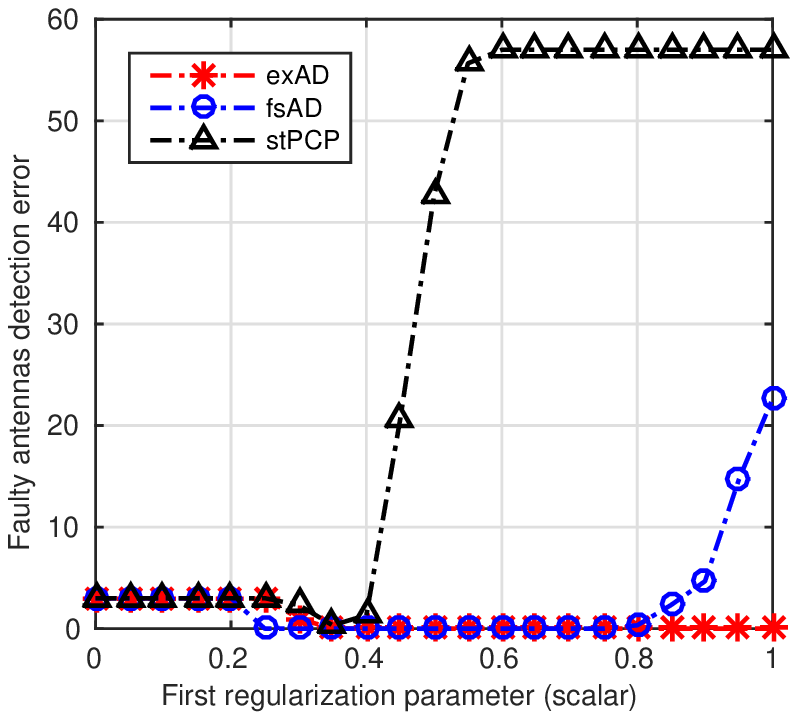}\label{fig: figure 2b}}
\caption{Comparison of the channel estimation methods for different values of $\alpha\in[0:0.05:1]$ with $\{\tau_1,\nu_1,\lambda_1\}=\alpha\|\Z\|_{2}$, $\{\tau_2,\nu_2,\lambda_2\}=0.5\|\Z\|_{\minfnorm}$, $M=60$, $S=\lfloor0.05M\rfloor$, $K=10$, $P=20$ and $L=10$.}
\label{fig: experiment 2}
\end{figure*}

Our proposed algorithms (exAD, fsAD) and the stPCP algorithm essentially solve a matrix decomposition problem based on $\Z$. The selection of each pair of regularization parameters follows the approach in \cite{parikh2013proximal} as below. From \eqref{eqn: SDP atomic norm}, \eqref{eqn: SDP nuclear norm} and \eqref{eqn: bound on new atomic norm}, we have
\begin{align}
\|\Z\|_{\mA_N}^{*}\leq\|\Z\|_{\mA}^{*}\leq\|\Z\|_{2},
\end{align}
where $\|\Z\|_{2}$ is the dual norm of $\|\Z\|_{*}$. The dual norm of $\|\Z\|_{\mnorm}$ is $\|\Z\|_{\minfnorm}:=\max_{i}\|\Z{(i,:)}\|_2$. The first and second regularization parameters of each algorithm are respectively upper bounded by $\|\Z\|_{2}$ and $\|\Z\|_{\minfnorm}$, above which the optimal values of $\H$ and $\W$ are zero. Hence, the regularization parameters are set to a down scaled version of $\|\Z\|_{2}$ and $\|\Z\|_{\minfnorm}$. We use $N=MK$ for the fast algorithm.

In the first simulation, we set the number of BS antennas $M=120$, the number of faulty antennas $S=\lfloor0.05M\rfloor$, number of paths $P=20$, length of pilot sequences $L=10$, the number of users $K=10$ and SNR$=10$ dB. To compare the performances of different regularization algorithms, we vary their first regularization parameters (i.e. $\tau_1$, $\nu_1$ and $\lambda_1$) at different scaling levels and keep the second regularization parameter fixed. The reason is that the first regularization parameters are associated with the penalties on the corresponding norms of the estimated channel matrix, which are different for each algorithm. On the other hand, the norms of the distortion matrix ($\|\W\|_{\mnorm}$) are the same for all three algorithms. We vary the scalar $\alpha=[0:0.05:1]$, and set $\{\tau_1,\nu_1,\lambda_1\}=\alpha\|\Z\|_{2}$ (exAD, fsAD and stPCP), $\{\tau_2,\nu_2,\lambda_2\}=0.5\|\Z\|_{\minfnorm}$. For each scalar and each algorithm, $10$ iterations are run to obtain the average channel estimation error and faulty antenna detection error.

Fig. \ref{fig: figure 1a} shows the channel estimation error for different methods at different scaling levels of the first regularization parameter\footnote{The $x$-axis denotes the scalar $\alpha$, while the actual value of the first regularization parameter is $\alpha\|\Z\|_{2}$.}. Obviously, the performances of LS, LS-SLS and MMSE based methods are independent of the value of the first regularization parameter. We note that the LS based method converges to the LS-SLS based one, while the MMSE method provides a better performance due to its extra assumption on knowing $\A$. For the regularization algorithms, the first regularization parameter traces out a path of solutions. The original channel matrices are overestimated when the scalar is too small, and over-shrunken when the scalar is too large. At scaling levels $\alpha=\{0.3, 0.2, 0.5\}$, the exAD, fsAD and stPCP respectively achieve the smallest channel estimation error. Although the MMSE method utilizes extra knowledge on the matrix $\A$, it is interesting to note that our proposed algorithms are comparable to the MMSE method. In addition, it indicates that both of them provide better channel estimation performance than the stPCP algorithm. Furthermore, the smallest channel estimation error given by the fast algorithm (fsAD) is comparable with that from the original algorithm (exAD), which is consistent with our theoretical results in \eqref{eqn: bound on new atomic norm}. In addition to channel estimation, the regularization algorithms simultaneously provide estimation on the locations of faulty antennas at the BS. Fig. \ref{fig: figure 1b} shows their performances on detecting faulty antennas. It indicates that the algorithms can faithfully detect faulty antennas when the scalar is within certain ranges.

For the second simulation, the setting is the same as the first one but the number of antennas at the BS reduced to $M=60$. As shown in Fig. \ref{fig: figure 2a}, the scaling levels at which the regularization algorithms achieve smallest channel estimation error are slightly changed to $\alpha=\{0.45, 0.3, 0.4\}$, which are rather stable considering the large reduction on $M$. Also, Fig. \ref{fig: figure 2b} indicates that the ranges within which each algorithm can faithfully detect faulty antennas cover the corresponding points $\{0.45, 0.3, 0.4\}$.

\begin{figure}[!t]
\centering
\includegraphics[width=3.5in]{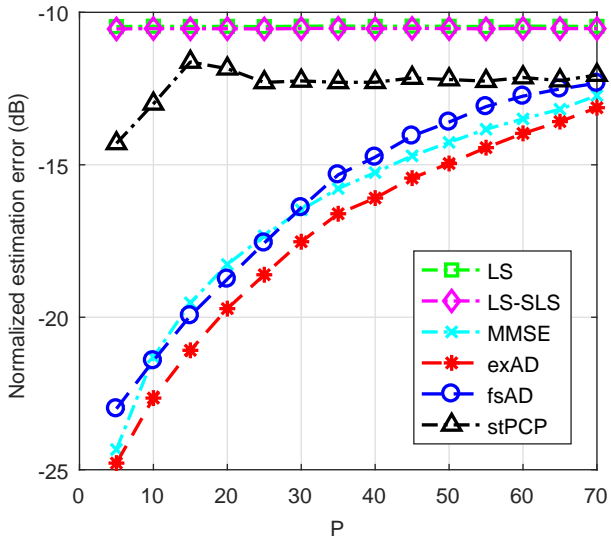}\label{fig: figure3}
\caption{Comparison of the channel estimation methods for different values of $P$ with $\alpha=\{0.3, 0.2, 0.5\}$, $\{\tau_1,\nu_1,\lambda_1\}=\alpha\|\Z\|_{2}$, $\{\tau_2,\nu_2,\lambda_2\}=0.5\|\Z\|_{\minfnorm}$, $M=120$, $S=\lfloor0.05M\rfloor$, $K=10$, $P=20$ and $L=10$.}
\label{fig: experiment 3}
\end{figure}

\emph{The impact of the number of AoAs.} In this simulation, we investigate the channel estimation performances of different algorithms with respect to different number of directions $P$. The setting is the same as that of the first simulation, but the scaling levels of the first regularization parameters are fixed at $\alpha=\{0.3, 0.2, 0.5\}$. As shown in Fig. \ref{fig: experiment 3}, the performances of LS based and LS-SLS based methods are independent of the value of $P$. On the other hand, the performances of the rest algorithms degrade as $P$ increases.

\begin{figure}[!t]
\centering
\includegraphics[width=3.5in]{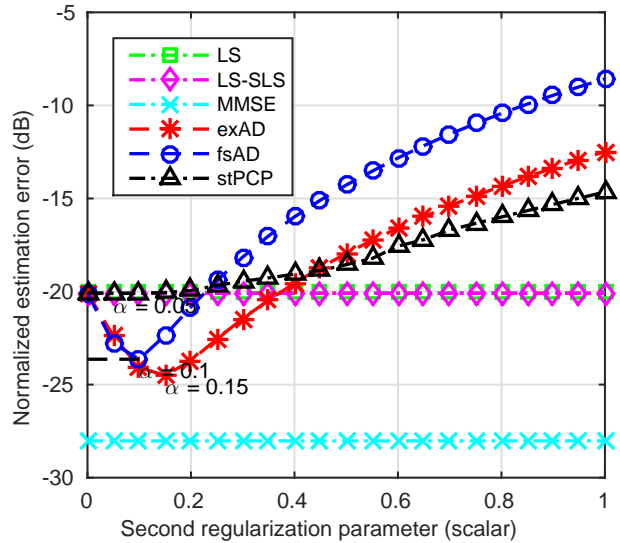}\label{fig: figure4}
\caption{Comparison of the channel estimation methods for different values of $\alpha\in[0:0.05:1]$ with $\{\tau_1,\nu_1,\lambda_1\}=\alpha\|\Z\|_{2}$, $\{\tau_2,\nu_2,\lambda_2\}=0.5\|\Z\|_{\minfnorm}$, $M=120$, $S=0$, $K=10$, $P=20$ and $L=10$.}
\label{fig: experiment 4}
\end{figure}

\emph{The impact of the faulty antennas on MMSE.} In the above simulations, it is noted that the proposed algorithms can sometimes perform better than the MMSE estimation method. When the noise term is Gaussian distributed, using MMSE estimation is optimal \cite{kay1993fundamentals}. In the above simulations, the received signal is affected by both the additive noise $\mathbf{N}_0$ and the corruption noise $\mathbf{W}_0$ due to the faulty antennas. Therefore, even with the knowledge on the angular support $\A$, using MMSE estimation may not always provide optimal performance since the noise term $\mathbf{N}_0+\mathbf{W}_0$ is not Gaussian distributed. To further elaborate this point, simulation results considering zero faulty antenna are included here. In this case, the only noise term $\mathbf{N}_0$ is Gaussian distributed. The setting of this simulation is the same as the first one except for two things: the number of faulty antennas is zero, i.e. $S=0$, and the values of the second regularization parameters ($\tau_2$, $\nu_2$, $\lambda_2$) are set to infinity such that the estimated number of faulty antennas would always be zero. As shown in Fig. \ref{fig: experiment 4}, MMSE estimation exhibits the best performance, which is consistent with the theoretical understanding.

\begin{figure}[!t]
\centering
\includegraphics[width=3.5in]{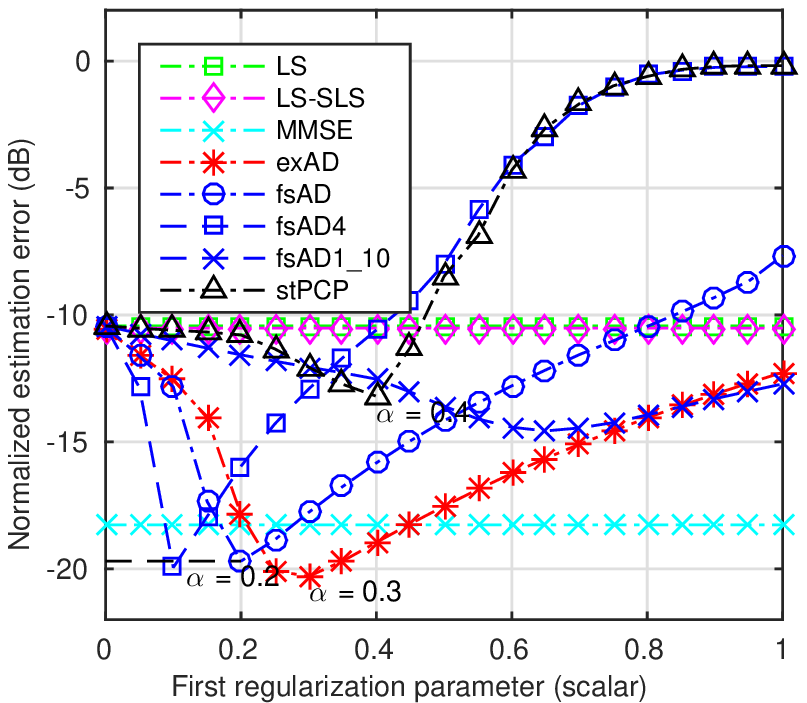}\label{fig: figure5}
\caption{Comparison of the channel estimation methods under random AoAs for different values of $\alpha\in[0:0.05:1]$ with $\{\tau_1,\nu_1,\lambda_1\}=\alpha\|\Z\|_{2}$, $\{\tau_2,\nu_2,\lambda_2\}=0.5\|\Z\|_{\minfnorm}$, $M=120$, $S=\lfloor0.05M\rfloor$, $K=10$, $P=20$ and $L=10$.}
\label{fig: experiment 5}
\end{figure}

\emph{The impact of random AoAs.} Next, we demonstrate the impact of the distribution of the AoAs on the performance. The simulation follows the settings of the first one except that the AoAs are chosen randomly in $[-\pi/2, \dots, \pi/2]$. Three different resolution settings are selected for the fsAD algorithm: fsAD1\text{\_}10 with $N=MK/10$, fsAD with $N=MK$ and fsAD4 with $N=4MK$. As shown in Fig. \ref{fig: experiment 5}, fsAD ($N=MK$) is still able to provide a low normalized estimation error. For $N=4MK$, the performance is further improved and becomes very close to that of exAD. On the other hand, the normalized estimation error increases when $N=MK/10$, which indicates that the resolution of the grid (i.e. $N$) does influence the performance. The simulation results are consistent with the theoretical implications.

\begin{figure}[!t]
\centering
\includegraphics[width=3.5in]{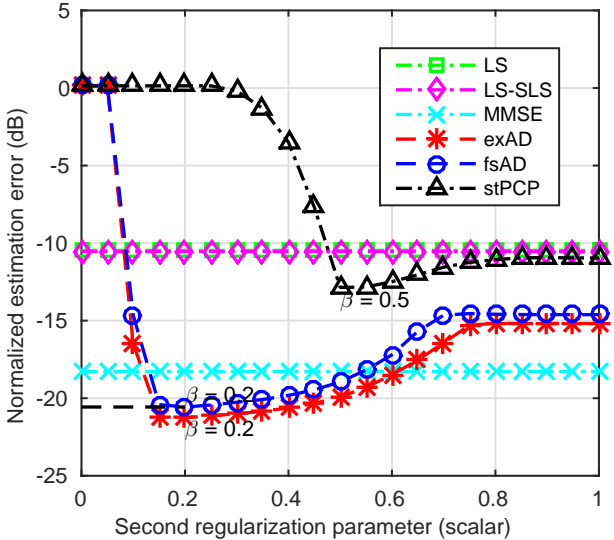}\label{fig: figure6}
\caption{Comparison of the channel estimation methods for different values of $\beta\in[0:0.05:1]$ with $\alpha=\{0.3, 0.2, 0.5\}$, $\{\tau_1,\nu_1,\lambda_1\}=\alpha\|\Z\|_{2}$, $\{\tau_2,\nu_2,\lambda_2\}=\beta\|\Z\|_{\minfnorm}$, $M=120$, $S=\lfloor0.05M\rfloor$, $K=10$, $P=20$ and $L=10$.}
\label{fig: experiment 6}
\end{figure}

\emph{The effect of the second regularization parameters.} In previous simulations, we vary the first regularization parameters (i.e. $\tau_1$, $\nu_1$ and $\lambda_1$) at different scaling levels and keep the second regularization parameters fixed. By doing so, we are able to demonstrate the effects of other factors (i.e., $M$, $P$, and AoAs). We include additional simulation results here to demonstrate the impact of the second regularization parameters on the performance. We fix the first regularization parameters and vary the second regularization parameters at different scaling levels. As shown in Fig. \ref{fig: experiment 6}, the performances of the proposed methods (exAD and fsAD) are relatively stable when the scalars of second regularization parameters are in the range of $(0.15, 0.55)$.

\emph{Signal detection performance} Lastly, we examine the signal detection performances associated with different channel estimation methods. In this simulation, we apply the same settings as in the first two but fix $\alpha=\{0.4, 0.25, 0.4\}$ and $\{\tau_2,\nu_2,\lambda_2\}=0.5\|\Z\|_{\minfnorm}$. We obtain the estimated channel matrices by different channel estimation methods for different numbers of BS antennas ranging from $60$ to $120$. Then, each estimated channel matrix is employed for signal detection over $10^5$ signal samples. For channel matrices estimated by LS or LS-SLS methods, the traditional ZF receiver is employed in the signal detection process due to its good performance at high SNR. Since our proposed method can estimate the channel matrix and locate the indices of faulty antennas simultaneously, the improved ZF receiver is applied for signal detection.

\begin{figure*}[!t]
\centering
\subfloat[Channel Estimation]{\includegraphics[width=3.5in]{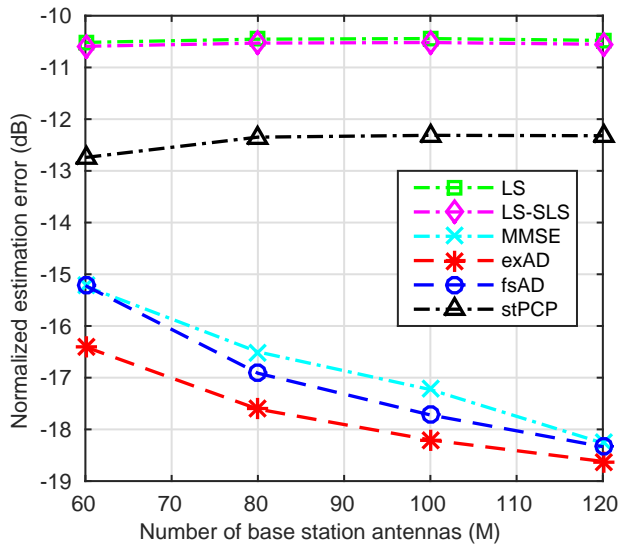}\label{fig: figure7a}}
\hfil
\subfloat[ZF Receiver with BPSK modulation]{\includegraphics[width=3.5in]{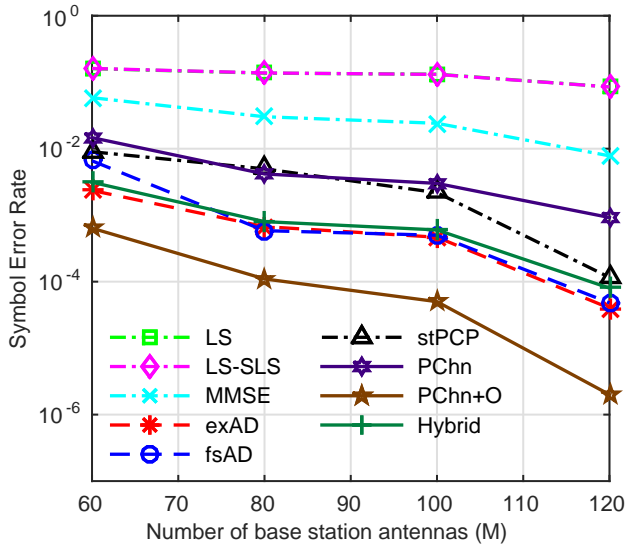}\label{fig: figure7b}}
\caption{Comparison of the signal detection performances associated with channel estimation methods for different values of BS antennas $M$ with $S=\lfloor0.05M\rfloor$, $K=10$, $P=20$ and $L=10$.}
\label{fig: experiment 7}
\end{figure*}

Fig. \ref{fig: figure7a} shows the channel estimation error for different methods. It can be seen that the proposed approaches achieve better performance. The symbol error rate for 8-PSK modulation based on different estimated channel matrices are shown in Fig. \ref{fig: figure7b}. The figure also presents two oracle performances: `PChn' represents the results obtained by assuming perfect knowledge on the channel matrix and using the traditional ZF receiver; `PChn+O' represents the results obtained by assuming perfect knowledge on both the channel matrix and the indices of faulty antennas and employing the improved ZF receiver. The performance gap between the two oracle curves verifies our asymptotic analysis on the benefit of knowing the indices of faulty antennas in Section \ref{sec: chn est with defective antennas}. In addition, the performance gap between the MMSE and our proposed algorithms further demonstrates such a benefit. It can be seen that our proposed methods provide better signal detection performance than other methods.

\section{Discussions}\label{sec: discussion}
\subsection{SDP characterizations of the algorithms}
Intuitively, the superior performance of our proposed method can be attributed to its better exploration on the intrinsic information, that is, the application of atomic norm instead of nuclear norm. Specifically, this can be seen through the SDP characterizations of algorithms \eqref{eqn: prox algo atomic norm} and \eqref{eqn: prox algo nuclear norm}.

For the atomic norm-based approach \eqref{eqn: prox algo atomic norm}, we have the following SDP characterization
\begin{align}\label{eqn: SDP atomic norm}
\underset{\a,\B,\H}{\argmin}&\ \frac{1}{2}\Tr(\mathcal{T}(\a))+\frac{1}{2}\Tr(\B)+\mu_1\|\H-\V_1^k\|_{F}^2\\
\text{s.t.}&\quad \begin{bmatrix}\mathcal{T}(\a) & \H \\
                                \H^*            & \B\end{bmatrix}\succeq 0\nonumber,
\end{align}
where $\mathcal{T}(\a)$ is the Hermitian Toeplitz matrix with vector $\a$ as its first column and $\mu_1$ depends only on $\tau_1$.

The SDP characterization of \eqref{eqn: prox algo nuclear norm} is given by
\begin{align}\label{eqn: SDP nuclear norm}
\underset{\P,\Q,\H}{\argmin}&\ \frac{1}{2}\Tr(\P)+\frac{1}{2}\Tr(\Q)+\mu_2\|\H-\V_1^k\|_{F}^2\\
\text{s.t.}&\quad \begin{bmatrix}\P   & \H \\
                                \H^* & \Q\end{bmatrix}\succeq 0\nonumber,
\end{align}
where $\mu_2$ depends only on $\lambda_1$. It can be observed that \eqref{eqn: SDP nuclear norm} is a relaxation of \eqref{eqn: SDP atomic norm} by dropping the Toeplitz constraint of the first diagonal block in the positive semi-definite (PSD) feasibility condition, which explains why our proposed atomic norm-based method outperforms the stable PCP-based one.

\subsection{Regularization parameters tuning in practice}
In practical applications, the proposed atomic norm denoising-based algorithms (exAD and fsAD) depend on unknown regularization parameters that need to be chosen carefully, which is known as ``parameter tuning''. For general parameterized semidefinite convex optimization problems, there have been some studies on performing the tuning step offline by computing an approximated regularization path \cite{d2007full,giesen2012regularization,mateos2012robust} and cross-validation algorithms \cite{arlot2010survey}. In this part, we propose a hybrid approach to choose the regularization parameters for the atomic norm denoising algorithms. The difficulty of choosing the regularization parameters in practice is the lack of a reference. In other words, for each pair of regularization parameters, the quality of the estimated channel matrix can not be verified without the knowledge on the true channel matrix. From the first two simulations, it is observed that the channel estimation results given by MMSE are close to those of the proposed algorithms (exAD and fsAD). Thus, our hybrid approach is as below.
\begin{enumerate}
  \item Assume that $\A$ is known, and estimate the channel matrix via MMSE.
  \item Use the MMSE estimated channel matrix as a reference, and test exAD/fsAD with different pairs of regularization parameters within in a certain range.
  \item Find the pair of regularization parameters such that the exAD/fsAD estimated channel matrix is close to the MMSE one.
  \item The estimated indices of faulty antennas given by that pair of regularization parameters can be obtained simultaneously.
  \item Do signal detection with the modified MRC/ZF receiver.
\end{enumerate}
The hybrid approach does not require a priori knowledge of the regularization parameters. It uses the MMSE estimation result as a reference and searches for the suitable pair of regularization parameters. In Fig. \ref{fig: figure7b}, the symbol error rate of the hybrid approach with exAD is depicted. It can be seen that the result is close to that given by the exAD method.
\section{Conclusion}\label{sec: conclusion}
In this paper, we have considered channel estimation and faulty antenna detection for massive MIMO systems. We have proven that the negative effect of faulty antennas on signal detection does not vanish even with unlimited number of BS antennas. To mitigate this effect, an atomic norm denoising-based approach has been proposed to simultaneously acquire the CSI and detect the locations of faulty antennas. In addition, we have proposed a fast algorithm which has been shown to be a good approximation of the original algorithm. It has been demonstrated that both the proposed approaches outperform existing ones including conventional linear estimators and a stable PCP-based method.

We envision several directions for future work. Firstly, analysis on the effect of faulty antennas on other receivers (e.g., MMSE) and possible improvements could be investigated. Secondly, our analysis and algorithm have considered the worst case scenario, i.e, no statistical knowledge on the channel or the distortion caused by faulty antennas is assumed. With more measurement results for massive MIMO systems with faulty antennas, it may provide a statistical model for the distortion noise, and hence help develop better algorithms. Lastly, the effect of faulty antennas could be analyzed together with other factors in practical massive MIMO systems, e.g., mutual coupling, element pattern, pilot contamination, and colored noise model.

\appendices
\section{Bound on the new atomic norm $\|\H\|_{\mA_{N}}$}\label{app: bound on new atomic norm}

By definition, we can write the dual norm of $\|\H\|_{\mA}$ as
\begin{align}
\|\V\|_{\mA}^{*}&=\sup_{\|\H\|_{\mA}\leq 1}\langle\V,\H\rangle\nonumber\\
                &=\sup_{\substack{f\in[0,1)\\ \|\u\|_2=1}}\left|\left\langle\u,\frac{1}{\sqrt{M}}\V^*\a(f)\right\rangle\right|\nonumber\\
                &=\sup_{f\in[0,1)}\left\|\frac{1}{\sqrt{M}}\V^*\a(f)\right\|_2\nonumber\\
                &=\sup_{f\in[0,1)}\sqrt{\sum_{k=1}^K|v_k(f)|^2},\label{eqn: dual norm}
\end{align}
where $\langle\V,\H\rangle:=\text{Re}(\text{Tr}(\V^*\H))$ and
\begin{align}
v_k(f)=\frac{1}{\sqrt{M}}\sum_{i=1}^M\V_{(i,k)}^*\mathrm{e}^{j2\pi f(i-1)}.
\end{align}

Similarly, the dual norm of $\|\H\|_{\mA_{N}}$ is given by
\begin{align}\label{eqn: dual norm with f}
\|\V\|_{\mA_{N}}^*=\sup_{f\in[(N-1)/N]}\sqrt{\sum_{k=1}^K|v_k(f)|^2},
\end{align}
where $[(N-1)/N]:=\{0,1/N,\dots,(N-1)/N\}$.

For $f_1,f_2\in[0,1)$, by Bernstein's theorem (Lemma \ref{lem: bernstein}) and a result in \cite[Appendix B]{yang2015gridless}, we have
\begin{align}
|v_k(f_1)|-|v_k(f_1)|\leq 2\pi M|f_1-f_2|\sup_{f\in[0,1)}|v_k(f)|
\end{align}

\begin{lemma}[\cite{schaeffer1941inequalities}]\label{lem: bernstein}
Let $q(z)$ be any polynomial of degree $n$ with complex coefficients and derivative $q'(z)$. Then,
\begin{align}
\sup_{|z|\leq 1}|q'(z)|\leq n\sup_{|z|\leq 1}|q(z)|.
\end{align}
\end{lemma}

Thus,
\begin{align}
&\sum_{k}|v_{k}(f_1)|^2-\sum_{k}|v_{k}(f_2)|^2\\
&\quad\leq\sum_{k}(|v_{k}(f_1)|+|v_{k}(f_2)|)2\pi M|f_1-f_2|\sup_{f\in[0,1)}|v_k(f)|\nonumber\\
                                                               &\quad\leq\sum_{k}4\pi M|f_1-f_2|\sup_{f\in[0,1)}|v_{k}(f)|^2\nonumber\\
                                                                 &\quad\leq4\pi MK |f_1-f_2|\sup_{f\in[0,1)}\sum_{k}|v_{k}(f)|^2\nonumber\\
                                                                  &\quad=4\pi MK |f_1-f_2| (\|\V\|_{\mA}^*)^2.\label{eqn: vf bound}
\end{align}

Note that for any $f_1\in[0,1)$, there is a $f_2\in[(N-1)/N]$ with $|f_1-f_2|\leq\frac{1}{2N}$, then \eqref{eqn: vf bound} gives
\begin{align*}
\sup_{f\in[0,1)}\sum_{k}|v_{k}(f)|^2\leq&\sup_{f\in[(N-1)/N]}\sum_{k}|v_{k}(f)|^2\\
&\quad +\frac{2\pi MK}{N}(\|\V\|_{\mA}^*)^2.
\end{align*}
By substituting \eqref{eqn: dual norm} and \eqref{eqn: dual norm with f}, it becomes
\begin{align*}
(\|\V\|_{\mA}^*)^2\leq(\|\V\|_{\mA_N}^*)^2+\frac{2\pi MK}{N}(\|\V\|_{\mA}^*)^2.
\end{align*}
Hence,
\begin{align}
\|\V\|_{\mA}^*\leq(1-\frac{2\pi MK}{N})^{-\frac{1}{2}}\|\V\|_{\mA_N}^*.
\end{align}
From \eqref{eqn: dual norm} and \eqref{eqn: dual norm with f}, we have
\begin{align}
\|\V\|_{\mA_N}^*\leq\|\V\|_{\mA}^*.
\end{align}
Thus,
\begin{align}
\|\V\|_{\mA_N}^*\leq\|\V\|_{\mA}^*\leq(1-\frac{2\pi MK}{N})^{-\frac{1}{2}}\|\V\|_{\mA_N}^*,
\end{align}
and
\begin{align}
(1-\frac{2\pi MK}{N})^{\frac{1}{2}}\|\H\|_{\mA_N}\leq\|\H\|_{\mA}\leq\|\V\|_{\mA_N},
\end{align}
which completes the proof.

We note that the above techniques of bounding the dual norm associated with frequencies defined on uniform grids have been used in \cite{bhaskar2011atomic,li2014off,yang2015gridless}. Our proof generalizes the line spectral estimation in \cite{bhaskar2011atomic,yang2015gridless} into a mutiple measurement vectors model and improves the lower bound in \cite{li2014off}.


\bibliographystyle{IEEEtran}
\bibliography{refs}

\end{document}